\documentclass[a4paper, 11pt]{article}

\usepackage[utf8]{inputenc}
\usepackage{graphicx} 
\usepackage{amsmath,amssymb}

\usepackage{graphicx}        
\usepackage{multicol}        
\usepackage[bottom]{footmisc}

\usepackage{newtxtext}       %
\usepackage[varvw]{newtxmath}       

\usepackage[frozencache,cachedir=.]{minted}
\usepackage{caption} 
\usepackage{subcaption}
\usepackage{float}
\usepackage{url}
\usepackage[sort&compress,numbers]{natbib}

\title{Software for Massively Parallel Quantum Computing}
\author{Thien Nguyen$^1$, Daanish Arya$^1$, Marcus Doherty$^1$, Nils Herrmann$^1$, \\Johannes Kuhlmann$^1$, Florian Preis$^1$, Pat Scott$^1$, and Simon Yin$^1$}
\date{%
    $^1$Quantum Brilliance Pty Ltd, The Australian National University, Australia%
}

\begin{document}

%
%
\maketitle

\abstract{Quantum computing has the potential to offer substantial computational advantages over conventional computing. Recent advances in quantum computing hardware and algorithms have enabled a class of classically parallel quantum workloads, whereby individual quantum circuits can execute independently on many quantum processing units. Here, we present the full-stack software framework developed at Quantum Brilliance to enable multi-modal parallelism for hybrid quantum workloads. Our software provides the capability to distribute quantum workloads across multiple quantum accelerators hosted by nodes of a locally-networked cluster, via the industry-standard MPI (Message Passing Interface) protocol, or to distribute workloads across a large number of cloud-hosted quantum accelerators.}

\section{Introduction}
\label{sec:Introduction}

Quantum computing has the potential to solve previously intractable problems, including the factoring of large numbers~\cite{shor1994algorithms}, efficiently discovering optima in large search spaces~\cite{farhi2014quantum}, modelling and simulation of quantum mechanical systems~\cite{aspuru2005simulated, georgescu2014quantum}, and solving large systems of equations~\cite{harrow2009quantum, wiebe2012quantum, berry2014high, lloyd2014quantum}.  Current quantum computers are however unable to outperform their classical counterparts. This is for a number of reasons:\begin{itemize}
\item quantum algorithms are not yet as developed as classical ones;
\item quantum bits, or qubits, are subject to environmental disturbance (so-called `decoherence'), and thus exhibit high error rates;
\item the need to retain coherence throughout a calculation means that there is effectively no intermediate memory (such as processor cache or RAM) compatible with quantum algorithms at present.
\end{itemize}
To harvest computational power from these imperfect qubits, a typical design pattern for near-term quantum programs is to combine quantum and classical components into a hybrid, heterogeneous workflow. In this hybrid execution model, a high degree of parallelism emerges, whereby the computational workload can be distributed across many quantum processing units (QPUs), with only classical data exchanged between them. We refer to this as applying `classical parallelism' to the quantum workload. 

This model of quantum computation requires a software framework designed from the ground up to take maximal advantage of existing parallelisation schemes in high-performance computing (HPC).  These include threading, message passing and offload to classical accelerators such as GPUs. The QB Software Development Kit (SDK) is specifically designed to exploit these classical parallelisation schemes, not just in its simulation of quantum hardware, but also in its execution of hybrid classical-quantum and pure quantum algorithms on real quantum hardware.

In Section \ref{sec:Background}, we provide more background on hybrid quantum computing and the possibilities it offers for parallelisation, particularly those arising from compact form-factor accelerators such as those developed by Quantum Brilliance (QB). We then describe the QB SDK (Section \ref{sec:sdk-overview}), illustrate its parallelisation capabilities (Section \ref{sec:Implementation}), and conclude (Section \ref{sec:Summary}). 

\section{Background}
\label{sec:Background}

\subsection{Quantum Brilliance Hardware}

Quantum Brilliance builds quantum accelerators based on nitrogen-vacancy (NV) centers in diamond. The remarkable thermal and electrical properties of diamond~\cite{chen2020optimisation, doherty2021quantum} allow for long coherence times of the qubits at room temperature and standard air pressure.  Individual qubits are realized by two of the spin states of a single nitrogen nucleus. The electron contained in each NV center acts as bus for initialization and readout of the nitrogen spin qubit, as well as a mediator for interactions between multiple qubits. The nuclear spin is controlled by radio frequency (RF) fields, while the electron can be manipulated by microwave (MW) fields. The readout and initialization is achieved by pumping the electronic state with green laser light with a wavelength of 532\,nm.

Operation at room temperature requires only off-the-shelf control electronics and optics. Such QPUs do not require any of the usual bulky or expensive infrastructure typically seen in other quantum technologies, such as liquid Helium cooling systems, ultra-high vacuum or very precise lasers. Together these offer a significant opportunity to miniaturize the QPU and corresponding accelerator card to a form factor comparable to that of modern graphic accelerators.  The current commercial realisation of this technology is a 2-qubit QPU, hosted in a 19" wide, rack-mountable 6 U factor chassis~\cite{nunez_2022}.

The emergence of compact form-factor quantum accelerators paves the way to many-QPU systems similar to conventional supercomputers, and to deployment in edge computing scenarios. As we explain in the following subsection, such accelerators may also deliver a decisive performance advantage over classical systems earlier than larger systems.

\subsection{Quantum Utility}
\label{sec:QuantumUtility}
Quantum devices can traverse a Hilbert space of dimension $2^N$ with only $N$ qubits. 
Therefore, they promise significant breakthroughs in the computational sciences \cite[e.g.][]{doherty2021quantum, Gill_2022, Rietsche_2022}. Future quantum computers may enable calculations deemed impossible on the best known classical hardware. Arguably, such a future may not be on the cards for the currently emerging devices of the NISQ (noisy intermediate-scale quantum) era~\cite{preskill2018quantum}. Fortunately, this does not imply that NISQ computing does not possess an industrial value. It merely requires a firmer definition of what constitutes a genuine \textit{quantum advantage}.

Any quantum device starts to become industrially useful once it outperforms a competitor classical device in terms of either: 
\begin{itemize}
    \item computing time, 
    \item accuracy, or
    \item power consumption.
\end{itemize}
Such quantum devices are therefore said to possess a certain degree of \textit{quantum advantage}. This advantage can be sub-categorised into the realms of \textit{quantum dominance} and \textit{quantum utility}. Quantum dominance is obtained once a device can perform calculations that would be otherwise impossible on \textit{any} existing classical computer. Quantum utility, on the other hand, is obtained once a quantum device outperforms a classical device of comparable size, weight, and cost. That performance could however be matched or exceeded by more capable existing classical hardware, or greater size, weight or cost. 

\begin{figure}[ht]
    \centering
    \includegraphics[scale=1]{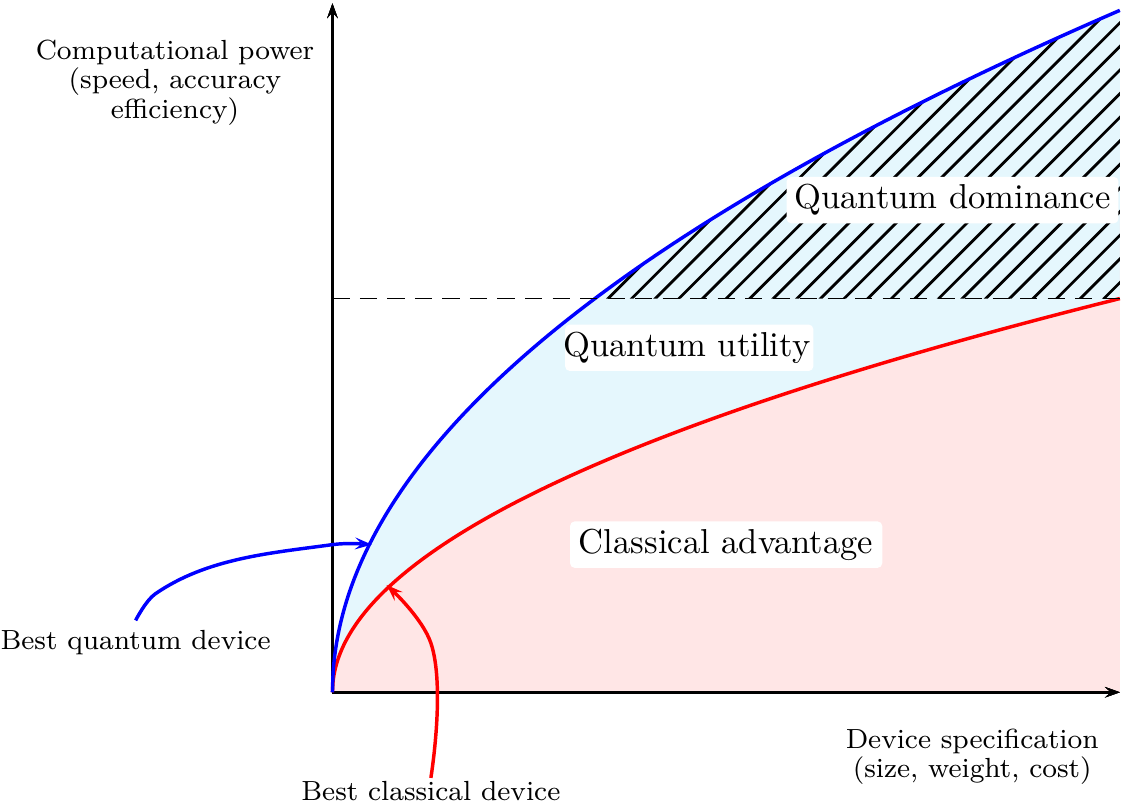}
    \caption{Schematic illustration of classical (light red) and quantum (light blue) advantage in a comparison of an abstract computational power and device specification. Here, computational power is a proxy for speed, accuracy, or efficiency, and device specification is a proxy for size, weight, and cost.  Note the splitting of the region of quantum advantage into quantum dominance (where no classical solution is possible) and quantum utility (where the same results can be obtained classically, but only by a higher-spec'd device than the device able to provide the quantum solution).}
    \label{fig:QuantumUtilityPlot}
\end{figure}

Fig. \ref{fig:QuantumUtilityPlot} provides a schematic visualization of the concept of quantum utility. Here, abstract computational power is plotted against an abstract `device specification' metric accounting for device size, weight and cost. The red and blue curves depict the performance of the best available classical devices, i.e.\ the boundaries of classical computing, and the best quantum devices, respectively. The light red area of classical advantage is separated from the light blue area of quantum advantage. The blue region splits further into the realm of quantum utility and quantum dominance, based on the availability or absence of larger, more costly, and more powerful classical devices.

While quantum dominance remains the ultimate goal of quantum computing, quantum utility may arguably be the more appropriate target for NISQ devices -- especially in a hybrid classical-quantum environment. Quantum utility marks a significant milestone for suppliers and customers of quantum technology alike, as its achievement is highly likely to usher in a phase of truly ubiquitous quantum computing.

\subsection{Hybrid Quantum Computing}

Several algorithms have been proposed to solve problems using pure quantum computing, but in the current NISQ era, quantum devices are still too small or too noisy to achieve non-trivial results with these techniques. This has been the motivation for the creation of hybrid algorithms, which leverage already-available classical resources to enhance the power of quantum computation, and vice versa. These hybrid algorithms primarily make use of variational quantum circuits with parameterised gates, and a classical optimizer that updates the parameters of the gates to achieve the desired result. Some examples of algorithms following this paradigm are given below. 

\subsubsection*{Variational Quantum Eigensolver}

Finding the ground state of a physical Hamiltonian is a problem of particular interest in the field of quantum chemistry. The Hamiltonian itself may have a structure that creates quantum correlations between an arbitrary number of its constituents. To solve this problem in a hybrid manner, one employs the \textit{variational quantum eigensolver (VQE)}~\cite{tilly2022variational}. VQE makes use of the natural quantum correlations within a quantum computer, through the creation of a circuit (or multiple circuits) that reflect the Pauli decomposition of the Hamiltonian under analysis. In addition to the variational ansatz, a state preparation circuit is applied (before) to rotate the system state into the desired basis, and a basis rotation circuit (after) is used to bring the state back into the measurement basis.

\subsubsection*{Quantum Approximate Optimization Algorithm}

The \textit{quantum approximate optimization algorithm (QAOA)}~\cite{farhi2014quantum} is a widely-used technique to solve graph problems on a quantum computer, in particular those reducible to a MaxCut or QUBO (quadratic unconstrained binary optimization) formulation. It does so by invoking the basic principle of adiabatic quantum computation (AQC), wherein a system is allowed to evolve under the influence of a pre-defined Hamiltonian for a set amount of time in order to arrive at a desired state. The variational ansatz itself is split into two parts: the cost, and the mixer. The cost Hamiltonian evolves the initial state according to the graph weights of the optimization problem, and the mixer Hamiltonian allows for traversal, or `mixing' between the allowable states of the optimizer. The classical optimizer is used to vary the rotation angles for the alternating variational ansatzes, which corresponds to varying Hamiltonian evolution time(s) within AQC. 

\subsubsection*{Quantum Machine Learning Algorithms}

Recent widespread adoption of machine learning has allowed scientists and engineers to gain intuition about using the same methods within a quantum context. Many \textit{quantum machine learning (QML)} algorithms have been created, with several quantum versions inspired by their classical counterparts: GANs $\to$ QGANs \cite{Zoufal2019QGAN}, support vector machines $\to$ QSVMs \cite{Rebentrost2014QSVM}, and convolutional neural networks $\to$ QCNNs \cite{Cong2019QCNN}. They each follow the hybrid paradigm of classically optimizing a variational circuit that is evaluated quantumly.  This is quite similar to the standard machine learning paradigm of optimizing linear transformations in the presence of non-linear activation functions. It has been suggested that in certain cases, QML algorithms can achieve similar accuracy to their classical counterparts, despite requiring less time~\cite{Abbas2021QNNPower}, or fewer data points~\cite{Caro2022GenQNN} for the training process. 

\subsection{Parallelism in Quantum Computing}

\begin{figure}[ht]
    \centering
    \includegraphics[width=\textwidth]{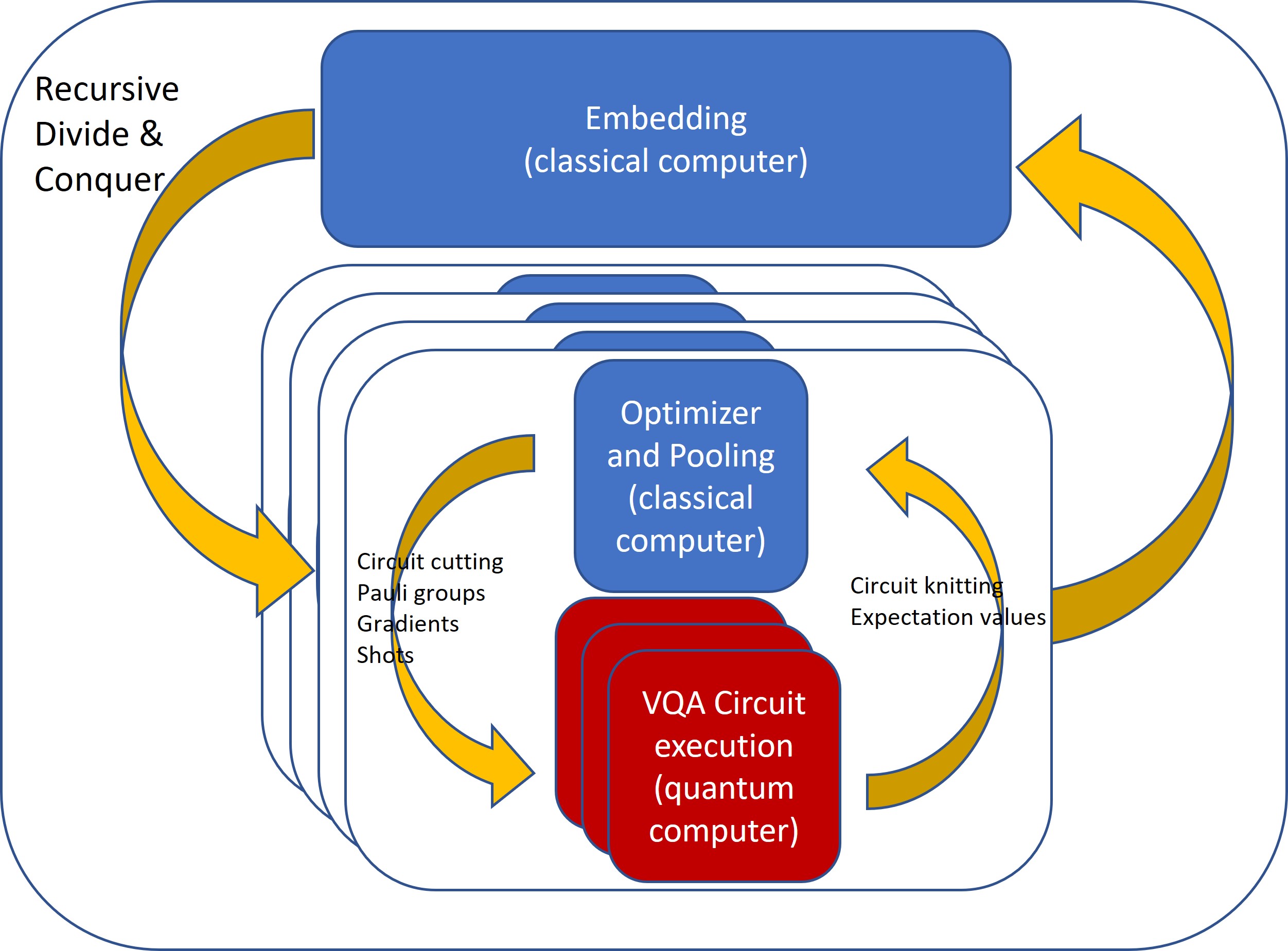}
    \caption{Illustration of nested self-consistent loops in an optimization problem involving the execution of a variational quantum algorithm (VQA) at the lowest layer. The outer layer recursively divides the full problem into sub-problems that can be solved in parallel, for example a QM/MM followed by a DFT/HF embedding scheme in quantum chemistry. Running the VQA itself within each sub-problem can be further parallelised across multiple QPUs.}
    \label{fig:EmbeddingAndParallelization}
\end{figure}

Many quantum-accelerated applications provide opportunities for classical parallelisation across multiple QPUs.  The different problem scales at which this can be realised are illustrated schematically in Fig \ref{fig:EmbeddingAndParallelization}.

At the highest level, applying a divide-and-conquer strategy to complex problems leads to an ensemble of quantum instructions~\cite{9537178} that can be trivially executed in parallel. A typical example of such a strategy is embedding methods in material sciences and chemistry. Consider a large molecule or a bath of many smaller molecules. In many cases the vast majority of the problem is sufficiently described by molecular mechanics (MM), i.e.\ the constituents are idealized as point charges and their interaction is governed by classical electrodynamics. In regions where quantum mechanical (QM) correlations cannot be neglected, e.g. where bonds may form or break, one can employ QM/MM schemes \cite{WARSHEL1976227}. In the so-called additive scheme the total energy of the system consists of the energy calculated for the QM region employing an appropriate QM method, while the energy for the MM region is simply the classical energy of interacting point charges. The interaction of both subsystems is accounted for in a third contribution in which the charge distribution of both subsystems is used as an input. These steps are repeated until the residual force fields fall below a given threshold, i.e. the full system equilibrates.

The QM regions can be tackled either directly on one or more QPUs, or by another nested approach whereby the computation on QPUs is embedded into an approximate QM method such as density functional theory (DFT), Hartree-Fock (HF) \cite{Rossmannek_2021}, or the coupled cluster (CC) approach. For instance, in the simplest cases one would freeze the core electrons treated with HF methods, restrict the active space in the electronic structure calculation performed on QPUs, and neglect the backreaction of the active electrons on the core.

Real world optimization problems in logistics, energy, financial services and manufacturing often require circuit widths that exceed the number of qubits likely to be available from quantum devices for the foreseeable future. Therefore, another promising approach to top-level parallelism is to divide the full problem into smaller sub-problems that can be solved on smaller quantum computers \cite[e.g.][]{https://doi.org/10.48550/arxiv.2101.07813, Pakhomchik:2022kyg}.

On the level of the QM computation, circuit knitting techniques \cite{Piveteau:2022ygd} such as entanglement forging \cite{PRXQuantum.3.010309} make it possible to partition larger quantum circuits into smaller ones.  The smaller circuits require less qubits and have smaller circuit depths than the original, but introduce significant overhead from the classical post-processing required to reconstruct the result of the original circuit from the results of the smaller ones.

At the lowest level, VQAs are particularly well suited for parallelization. For example, as discussed in~\cite{tilly2021variational}, the parallelization of VQEs applied to electronic structure calculations is a necessity to render these applications viable. 

The evaluation of the objective function often requires evaluation of many different circuits. For example, the Hamiltonian operator in electronic structure calculations consists of non-commuting terms that are accounted for by corresponding basis rotations in the quantum circuits. Depending on the Pauli grouping strategy, the number of distinct circuits scales linearly with the number spin-orbitals, albeit with a large pre-factor. Furthermore, in QML algorithms each data point in the training data set leads to a distinct circuit that contributes to the overall cost function. 

The classical optimization of the variational parameters requires further evaluations of this ensemble of circuits at different points in the parameter space within a single iteration step. 

Finally, in order to achieve a desired precision $\varepsilon$ (typically $\sim$$10^{-3}$ for chemistry), the execution of the same circuit must be repeated $1/\varepsilon^2$ times. These repetitions, often referred to as `shots', can be distributed over many QPUs.

\section{The Quantum Brilliance Software Development Kit (SDK)}

\label{sec:sdk-overview}
In order to realise the potential of hybrid algorithms and room-temperature quantum accelerators for quantum utility, one needs an appropriately parallelised quantum programming framework, specifically optimised for hybrid applications.  The Quantum Brilliance \emph{Software Development Kit} (SDK) is a full-stack software framework geared towards massive parallelism for accelerator-based hybrid quantum-classical workflows.

The QB SDK is designed for ease of integration into a range of performance-critical environments: HPC, cloud, GPU workstations and even embedded/edge systems.  The SDK's high-performance core and bundled simulators are implemented in C++.  It provides hybrid and pure-quantum programming interfaces in both C++ and Python, with an extensive language binding module for Python included. This allows users to access most of the underlying C++ features from Python, whilst simultaneously having access to Python's rich array of tools, such as NumPy for numerical programming and Matplotlib for plotting.

\begin{figure}[ht!]
\centering
\includegraphics[width=0.75\textwidth]{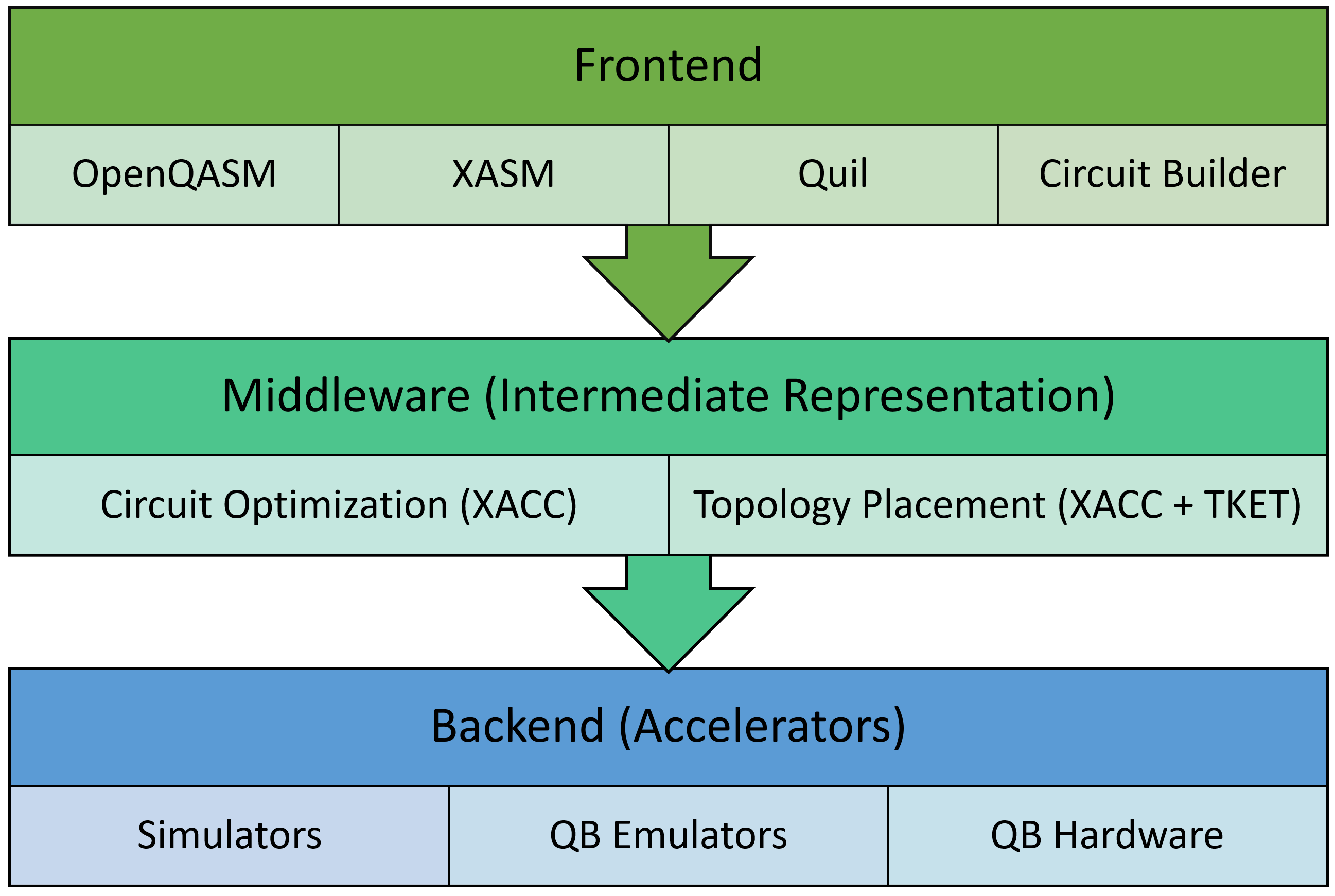}
\caption{The Quantum Brilliance Software Development Kit (SDK) provides a full-stack development environment for hybrid quantum-classical application development.}
\label{fig:sw-stack}
\end{figure}

Fig.~\ref{fig:sw-stack} depicts the high-level architecture of our SDK. This consists of (i) the frontend, which compiles various quantum programming inputs into an intermediate representation (IR); (ii) the middleware layer, which is a collection of software modules that transform and optimise quantum programs in IR format; and (iii) the backend, which converts IR into code directly executable on a specific simulator or actual quantum chips.

To ensure high performance and HPC compatibility, the current version of the SDK leverages the quantum intermediate representation (IR) of XACC \cite{xacc}, a modular and extensible system-level software infrastructure for heterogeneous programming. Here, we detail the functionality of the different layers as shown in Fig.~\ref{fig:sw-stack}, focusing on the extensions and modifications that we have made to XACC targeting our NV quantum devices and hybrid quantum-classical workflows. 

At the top of the stack, the QB SDK provides an extensible collection of frontend modules capable of parsing commonly-used quantum assembly dialects, such as IBM's OpenQASM~\cite{cross2017open} and Rigetti's Quil~\cite{smith2016practical}, based on XACC's \texttt{Compiler} interface for quantum compilation. We inject additional so-called \emph{intrinsics}, i.e.\ definitions of native quantum gates, pertinent to NV-based quantum computation, into the source assembly code directly, using e.g.\ the \texttt{include} directive of OpenQASM.

One notable feature of the SDK is the ability it offers for the user to programmatically construct quantum circuits via a utility called the \texttt{CircuitBuilder}. This allows users to write a program in high-level classical languages, such as C++ or Python, that describes how to build the quantum program. From this high-level program, the SDK constructs the relevant quantum kernel(s).  Examples of using the circuit builder utility to construct a simple two-qubit Bell-state circuit are shown in Fig.~\ref{fig:circuit_builder}.

\begin{figure}[h]
\begin{minted}[frame=single,framesep=10pt]{python}
# Import QB SDK core module
import qb.core
# Create an empty circuit 
simple_circuit = qb.core.Circuit()
# Add Hadamard gate on qubit 0
simple_circuit.h(0)
# Add a CNOT gate between qubit 0 and 1
simple_circuit.cnot(0, 1)
# Measure all qubits in the circuit
simple_circuit.measure_all()
\end{minted}
\begin{center}
(a) Python
\end{center}
\begin{minted}[frame=single,framesep=10pt]{C++}
// CircuitBuilder C++ header 
#include "circuit_builder.hpp"
// Create an empty circuit 
qb::CircuitBuilder simple_circuit;
// Add Hadamard gate on qubit 0
simple_circuit.H(0);
// Add a CNOT gate between qubit 0 and 1
simple_circuit.CNOT(0, 1);
// Measure all qubits in the circuit
simple_circuit.MeasureAll();
\end{minted}
\begin{center}
(b) C++
\end{center}
\caption{An example using the \texttt{CircuitBuilder} utility of the QB SDK.}
\label{fig:circuit_builder}
\end{figure}

As a convenience, we also provide a cloud-based user interface, whereby users have access to a Jupyter notebook instance with the latest SDK version installed and configured. Detailed information about this UI can be found in Ref.~\cite{qbos_orig}.   

Integral to the core of our SDK is the intermediate representation (IR) data structure, which encapsulates the semantics of quantum circuits. At the time of writing, the QB SDK relies on the XACC IR infrastructure.  This allows the SDK to make use of various composite IR modules in XACC that represent high-level quantum circuit templates such as for QAOA and QML, as well as XACC's middleware processing modules, including IR transformation passes for quantum circuit optimization and simplification. We have also implemented a module for interoperability with the open-source C++ library TKET~\cite{sivarajah2020t}, providing access to additional IR transformations such as noise-aware circuit mapping. 

The QB SDK has a large variety of backends for both production and experimental use.  These range from perfect (noise-free) quantum state simulation, to detailed emulation of quantum accelerators, to the ability to send compiled code to actual QB quantum chips. In particular, QB provides customers with an emulator module to enable accurate modeling of current and future hardware for algorithm and application development. With access to current QB hardware, users can also submit quantum circuits to the device via the SDK, with all steps from preparing the circuit for execution to sending it to the device handled by the SDK.

\section{Parallelism in the QB SDK}
\label{sec:Implementation}

A key design focus of the QB SDK is to support the quantum-accelerated high-performance computing vision whereby many quantum computing devices (QPUs) are integrated into conventional HPC data centers. To this end, we envision two modes of integration: (1) loosely-coupled or distributed integration and (2) integrated quantum-accelerated compute nodes. Specifically, the first refers to the case where the communication between classical HPC nodes and QPUs is mediated by conventional networks, thus HPC-QPU communication latency is not a big issue. In the latter scenario, QPUs would need to be integrated into the HPC infrastructure, including the high-speed, low-latency communication network (e.g., Infiniband), programming environment (e.g., MPI) and resource scheduler. We also note that scenario (1) encompasses both cloud-hosted (remote access over the Internet) and co-located (on-premises installation but connected via a LAN network, not integrated into the HPC network) use cases.   

To cover both of these modes in a flexible, efficient and performant manner, we have designed and implemented (1) an event-driven, shared-memory circuit execution interface to orchestrate a pool of loosely-coupled QPUs (such as those hosted on the cloud) and (2) a distributed-memory QPU virtualization scheme based on the gold-standard Message Passing Interface (MPI). The first model can take advantage of the availability of cloud-hosted QPUs by pooling them together while the latter can take advantage of a cluster of QPUs backed by a high-bandwidth, low-latency network. 

\subsection{Asynchronous QPU offloading}

\begin{figure}[tb]
\centering
\includegraphics[width=0.6\textwidth, angle=-90]{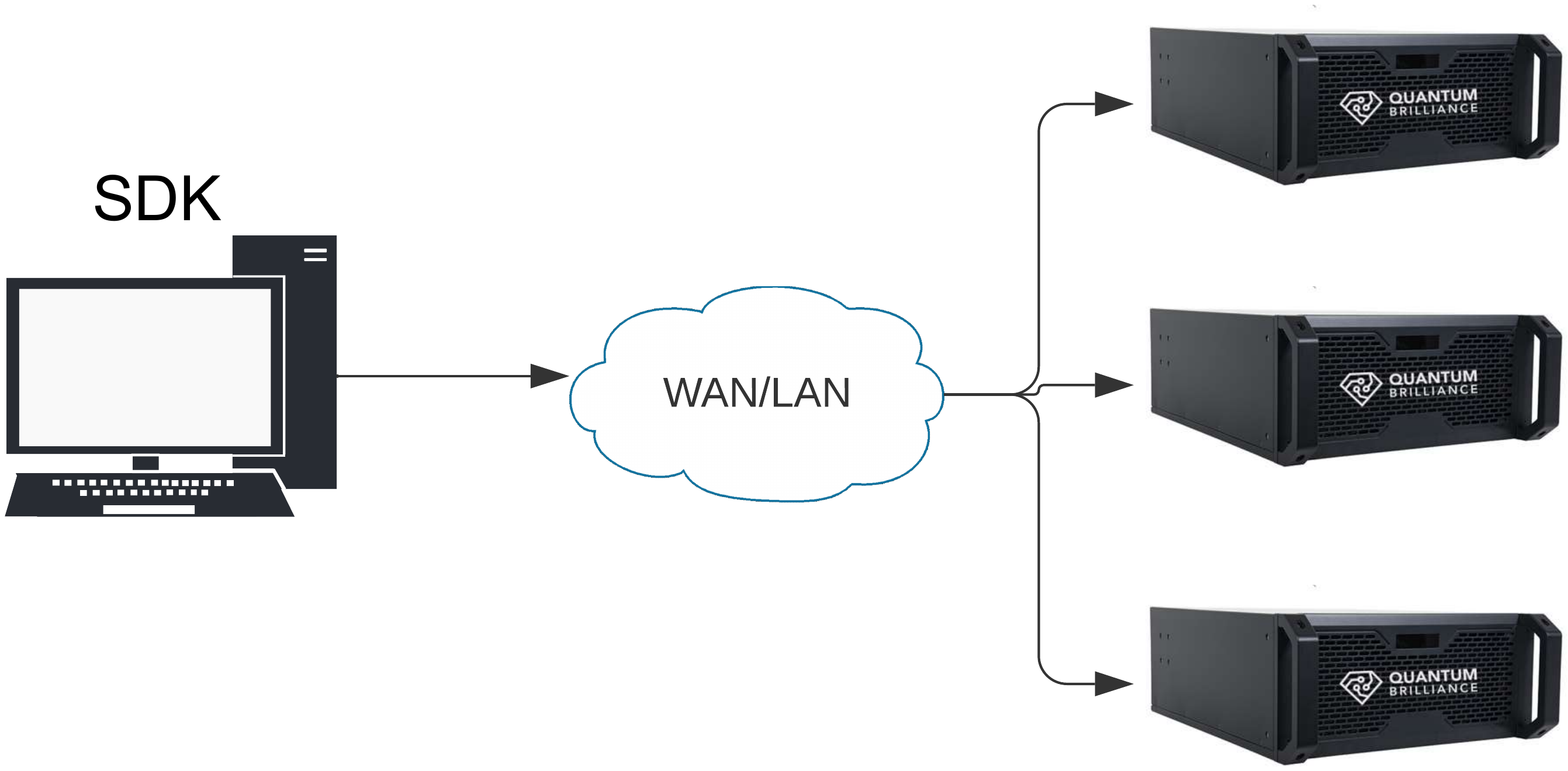}
\caption{Asynchronous QPU offloading: a single QB SDK thread (host program) distributing quantum tasks to multiple remotely-hosted QPUs. Asynchronous methods are used to query, wait for, and retrieve the results from the quantum accelerators.}
\label{fig:async_diagram}
\end{figure}

Most currently-available quantum computers are networked devices over private (LAN/Intranet) or public (WAN/Internet) networks. This means that they are typically communicated with via HTTP/REST protocols, as depicted in Fig.~\ref{fig:async_diagram}.

Our SDK features a high-level set of utilities for submitting concurrent parallel jobs to a set of remotely-hosted QPUs, using a non-blocking asynchronous offloading scheme for quantum operations. In this model, the execution of the quantum operations proceeds independently from the classical host program. Thus, the calling classical program can proceed to its subsequent instruction without waiting for completion of the quantum operations that were just offloaded. Usage of the results of the quantum operations requires explicit detection and handling by the calling program. 

\begin{figure}[tb]
\centering
\includegraphics[width=0.75\textwidth]{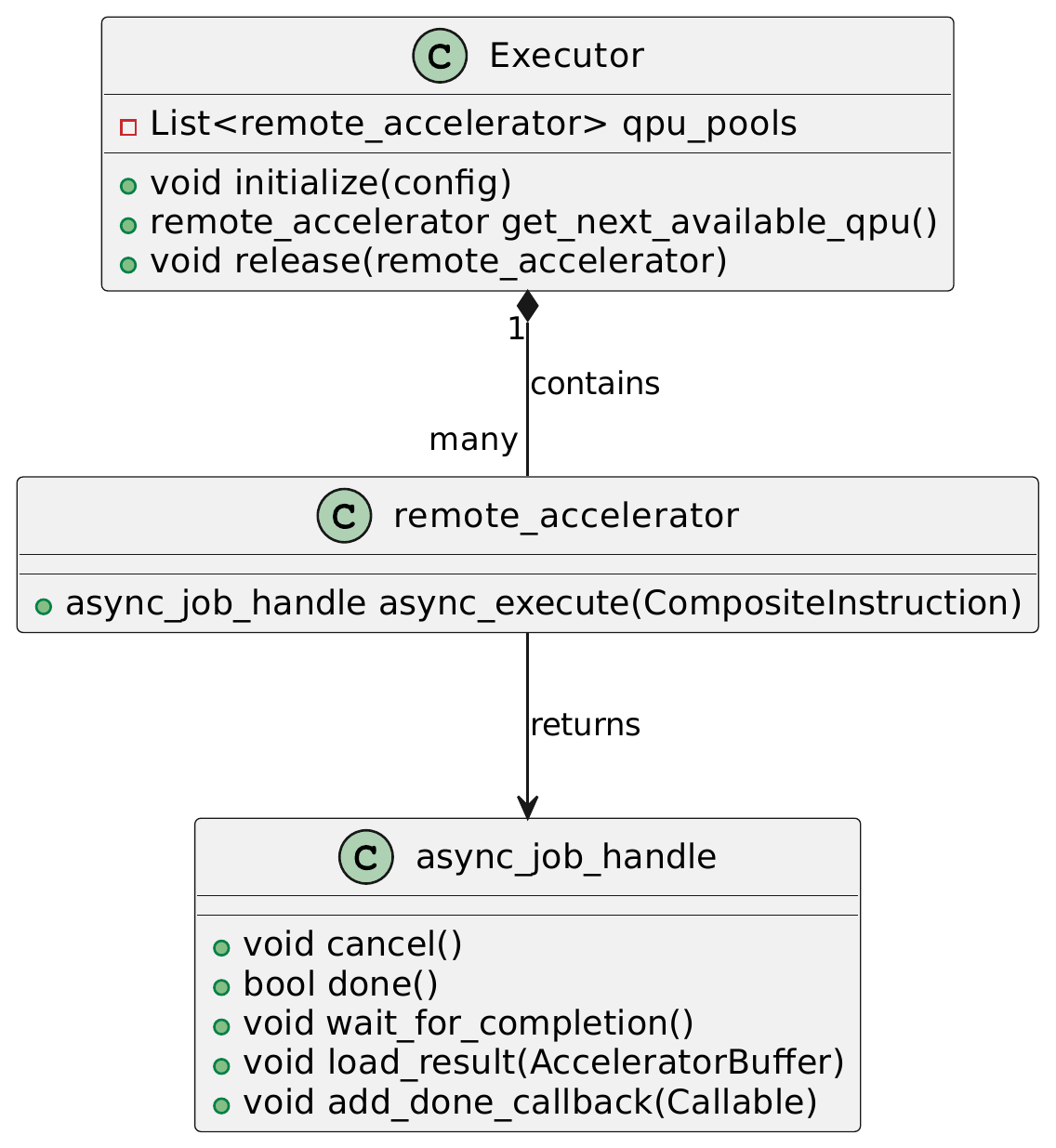}
\caption{Asynchronous QPU offloading architecture}
\label{fig:async_uml}
\end{figure}

Fig.~\ref{fig:async_uml} is a UML diagram describing the high-level constructs in the QB SDK that facilitate many-QPU parallelism via asynchronous offloading. The \texttt{remote\_}-\texttt{accelerator} represents a QPU target to which the SDK can offload tasks asynchronously via the \texttt{async\_}\texttt{execute} API. The quantum task is described by XACC's \texttt{CompositeInstruction} IR node, i.e.\ a quantum circuit.

The offloading API (\texttt{async\_execute}) is non-blocking.  It returns a handle, \texttt{async\_job\_handle}, from which the calling (host) program can check the status of the job and retrieve the results from the QPU. The asynchronous model tolerates a high level of variation in execution times between different accelerators. Lastly, a group of quantum accelerators can be bundled into an \texttt{Executor}, which is essentially a pool of remote quantum workers accessible in a round-robin fashion via the \texttt{get\_next\_available\_qpu} API. The \texttt{initialize} method provides the possiblity to configure an \texttt{Executor}'s properties, such as its number of QPUs and how to reach them.

\subsubsection*{Example}

The snippet in Listing \ref{lst:async_example} demonstrates how to use asynchronous offloading to submit many quantum tasks to a remote backend, such as available from AWS Braket.  

\begin{listing}[p]
\begin{minted}[frame=single,framesep=10pt, linenos, fontsize=\footnotesize]{python}
import numpy as np
import qb.core
my_session = qb.core.session()

# Helper to set up an executor consisting of up to 
# 32 AWS Braket SV1 backends
my_session.aws32sv1()

# Input parameterized quantum circuit in OpenQASM
my_session.instring = '''
__qpu__ void QBCIRCUIT(qreg q) {
    OPENQASM 2.0; 
    include "qelib1.inc";
    creg c[2];
    x q[1];
    ry(QBTHETA_0) q[0];
    cx q[1], q[0];
    measure q[0] -> c[0];
    measure q[1] -> c[1];
}'''

# Set up a parameter scan (100 data points)
scan_params = qb.MapND()
for theta in np.linspace(0, np.pi, 100):
    qb_theta = qb.ND()
    qb_theta[0] = theta
    scan_params.append(qb_theta)
# Set up the scan in the SDK job table 
my_session.theta[0] = scan_params
# List of all asynchronous job handles
job_handles = []
for idx in range(len(scan_params)):
    # Post all the jobs asynchronously 
    job_handle =  my_session.run_async(0, idx)
    job_handles.append(job_handle)

# At this point, all the jobs have been offloaded to 
# remote AWS Braket QPU(s)
# i.e., we can do other useful work in this thread
# ...

# The job handle can be used to check the status of a job
# e.g. check all the jobs are completed
all_done = all(handle.complete() for handle in job_handles)
\end{minted}
\caption{\label{lst:async_example} Example of asynchronous offload to remotely-hosted QPUs (AWS Braket).}
\end{listing}

Here we set up the \texttt{Executor} via an SDK utility function (\texttt{aws32sv1}) in line 7, which configures the executor to have up to 32 Braket SV1 simulator instances. The input circuit is provided as an OpenQASM source string (lines 10-20). To demonstrate the utility of asynchronous parallelism, we set up a parameterized quantum circuit pertinent to variational algorithms such as VQE or QAOA. A large number of quantum tasks are generated by scanning the parameter (lines 23-29). A 2D array represents the table of quantum jobs to run. The first index (row) refers to different quantum circuits, whereas the second index (column) refers to different run configurations, such as shot counts or circuit parameters as in this particular example.

With a large number of quantum jobs, we can use the Python \texttt{run\_async} API to enqueue a job from the table, ready to be sent to the executor. Here \texttt{run\_async} is just a high-level wrapper of the \texttt{get\_next\_available\_qpu} and \texttt{async\_execute} API functions shown in Fig.~\ref{fig:async_uml}, designed to post a quantum job to the backend in a non-blocking manner. As a result, the main thread is free to do other useful processing tasks after \texttt{run\_async} is invoked (as denoted in lines 37-40). The handle returned by each \texttt{run\_async} function call, of type \texttt{async\_job\_handle}, can be used to check the status of the job or retrieve the result when the job completes.

\subsection{Message Passing Interface}
\begin{figure}[tb]
\centering
\includegraphics[width=0.6\textwidth, angle=-90]{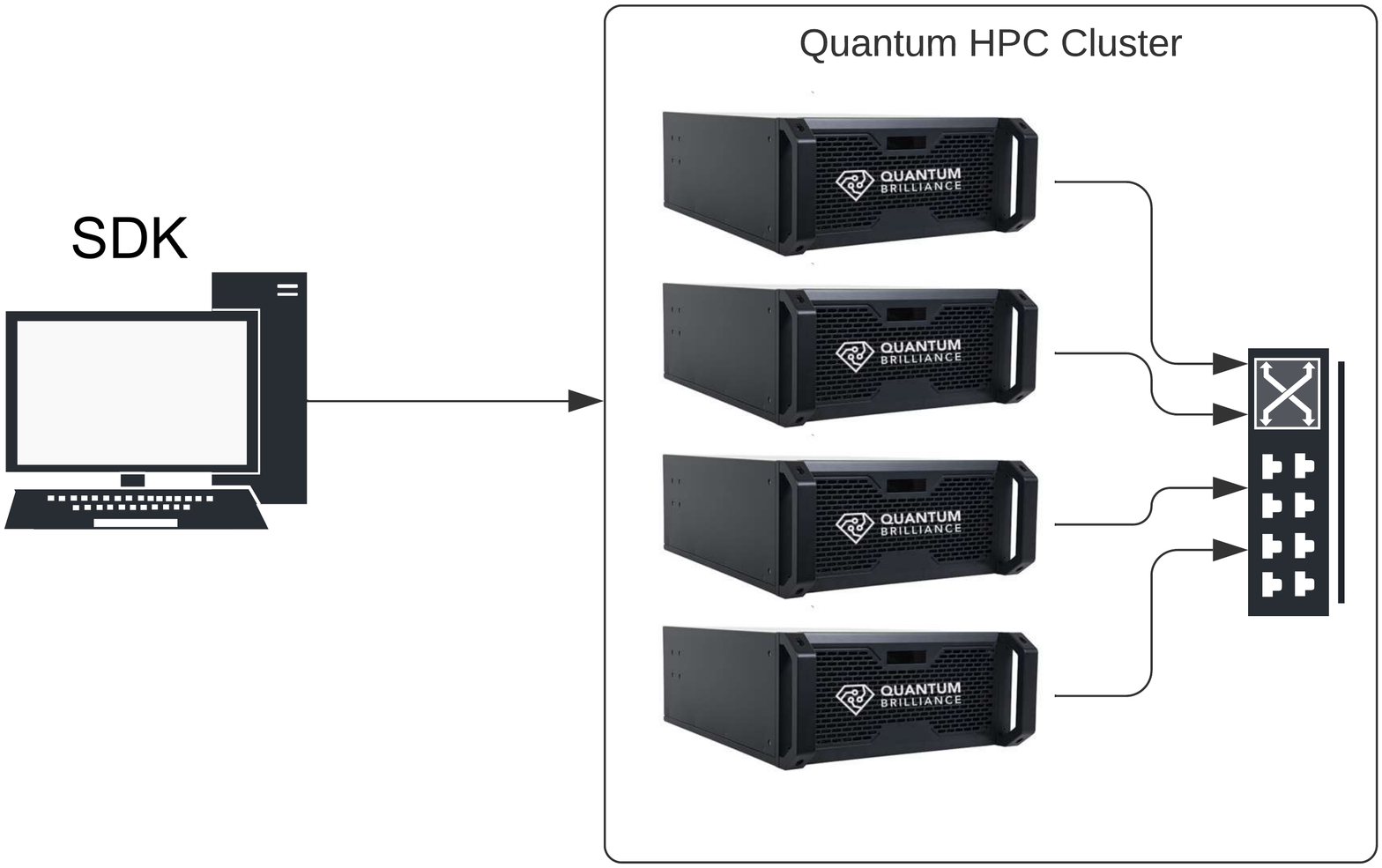}
\caption{MPI-based client-server architecture. The QB SDK UI (Jupyter notebook) is hosted in a conventional single-process context, connecting to a cluster of quantum-classical compute nodes. The cluster is backed by a high-speed network and MPI runtime.}
\label{fig:hpc_diagram}
\end{figure}

MPI is the most common parallelisation protocol used in high-performance computing (HPC). It allows for high bandwidth, low latency communication between compute nodes in a cluster. The QB SDK offers a client-server MPI architecture, where the user interface runs as a standalone client process that sends circuits to an MPI-enabled backend server for compute.  This design is depicted in Fig.~\ref{fig:hpc_diagram}. 

As was the case for its predecessor \cite{qbos_orig}, the SDK offers a user-friendly Jupyter notebook interface. The client-server model allows the UI to be executed as a conventional single-process program, running on e.g.\ a cluster login node or the cloud. The server code running on the quantum cluster uses MPI to distribute quantum tasks across all available nodes and gather the results. The rank 0 process of the server code handles the communication with the SDK UI client.

The server's core functionality is currently provided by an MPI-parallel wrapper around standard XACC accelerator backends (\texttt{HPCVirtDecorator}, derived from \texttt{AcceleratorDecorator}). The wrapper introduces pre- and post-processing steps. This provides simple parallelisation of serial and threaded accelerator backends. An example handling many small circuits utilizing the \texttt{HPCVirtDecorator} is given in Listing~\ref{listing:hpcDecorator}. The input parameter \texttt{params} contains descriptions for the ansatz and Pauli terms, and thus describes the quantum problem. \texttt{nWorker} and \texttt{nThreads} set the number of MPI processes and the number of threads per MPI process, respectively. Each MPI process handles exactly one accelerator backend, which may be further parallelized by threading. A simulator backend that makes native use of MPI can then also be orchestrated by the \texttt{HPCVirtDecorator}, using a multi-level MPI hierarchy. This makes it possible to spread multiple quantum circuits, each requiring more than one compute node, over a large compute cluster.

The \texttt{HPCVirtDecorator} can also be used to orchestrate the execution of quantum problems on analog QPUs. Here a circuit must fit into a single QPU, but multiple QPUs can work on the large list of Pauli terms or gradient computation points in parallel. 

\begin{listing}[p]
\begin{minted}[frame=single,framesep=10pt, linenos, fontsize=\tiny]{cpp}
// Here VQE is called with a decorated accelerator. The decorator adds pre- and 
// post-processing around the actual accelerator execution. This is used to introduce
// MPI parallelism, i.e partitioning and distribution of the vector of instructions 
// (base curcuit + Pauli terms) and return of the results. 
// Number of MPI processes and threads can be chosen as needed.
void xaccOptimizeDecorated(vqe::Params& params, int nWorker, int nThreads)
{ 
  // 1 of 4: ansatz from XACC qasm string
  std::shared_ptr<xacc::CompositeInstruction> ansatz;
  if (params.ansatz) 
  {
    // ansatz provided
    ansatz = params.ansatz;
  }
  else
  {
    // circuit string provided
    xacc::qasm(params.circuitString); 
    ansatz = xacc::getCompiled("ansatz");
  }
  
  // 2 of 4: observable from string
  std::shared_ptr<xacc::Observable> observable =
   std::make_shared<xacc::quantum::PauliOperator>();
  observable->fromString(params.pauliString);
  
  // 3 of 4: accelerator - qpp: "vqe-mode"=true is non-stochastic
  xacc::HeterogeneousMap accParams{{"n-virtual-qpus", nWorker}, 
                                   {"vqe-mode", params.isDeterministic}, 
                                   {"shots", params.nShots}, 
                                   {"threads", nThreads}};
                                 
  // get and wrap accelerator with hpc-decorator to introduce MPI parallelism
  auto accelerator = xacc::getAccelerator("qpp", accParams);
  accelerator = xacc::getAcceleratorDecorator("hpc-virtualization", 
                                              accelerator,
                                              accParams);
  
  // 4 of 4: optimiser
  auto optimizer = xacc::getOptimizer("nlopt");
  optimizer->setOptions({{"initial-parameters", params.theta},
                         {"nlopt-optimizer",    "cobyla"},
                         {"nlopt-maxeval",      params.maxIters},
                         {"nlopt-ftol",         params.tolerance}});
  
  // instantiate XACC VQE
  auto vqe = xacc::getAlgorithm("vqe");
  vqe->initialize({{"ansatz", ansatz},
                   {"accelerator", accelerator},
                   {"observable", observable},
                   {"optimizer", optimizer}});
  
  // Allocate some qubits and execute
  auto buffer = xacc::qalloc(params.nQubits);
  vqe->execute(buffer);
  
  // read out buffer 
  params.energies       = (*buffer)["params-energy" ].as<std::vector<double>>();
  params.theta          = (*buffer)["opt-params"    ].as<std::vector<double>>();
  params.optimalValue   = (*buffer)["opt-val"       ].as<double>();
}
\end{minted}
\caption{Example of MPI-based VQE, parallelised using an HPC decorator.}
\label{listing:hpcDecorator}
\end{listing}

\subsubsection*{Example}

Here, demonstrate the utility and performance of the QB SDK using a simulated quantum HPC cluster, where QPUs are replaced by classical numerical simulators running on conventional compute nodes. As quantum hardware continues to advance in terms of its quantum utility, the same software can be used to orchestrate a comparable large-scale deployment of quantum accelerators. 

In this example, we look at the problem of computing the expectation value of a complex Hamiltonian operator for a quantum state. This procedure is pertinent in many near-term quantum applications, such as the VQE algorithm. Specifically, given a Hamiltonian expressed in terms of Pauli operators,
\begin{equation}
    H = \sum_i \left(\bigotimes_j^{n_{\rm qubits}} \sigma_j^{(i)}\right), \sigma_j^{(i)} \in \{I, X, Y, Z\}
\end{equation}
the expectation value (energy) of a quantum state ($\langle \psi | H  | \psi \rangle$) can be computed by evaluating each term in the sum independently.

We look at the example of an $H_8$ molecule, consisting of four loosley-bonded $H_2$ molecules. The PySCF~\cite{sun2018pyscf} second-quantized Hamiltonian, which contains 1-particle and 2-particle integrals, can be converted into the above Pauli form using the Jordan-Wigner transformation.

\begin{figure}[tb]
\centering
\includegraphics[width=0.85\textwidth]{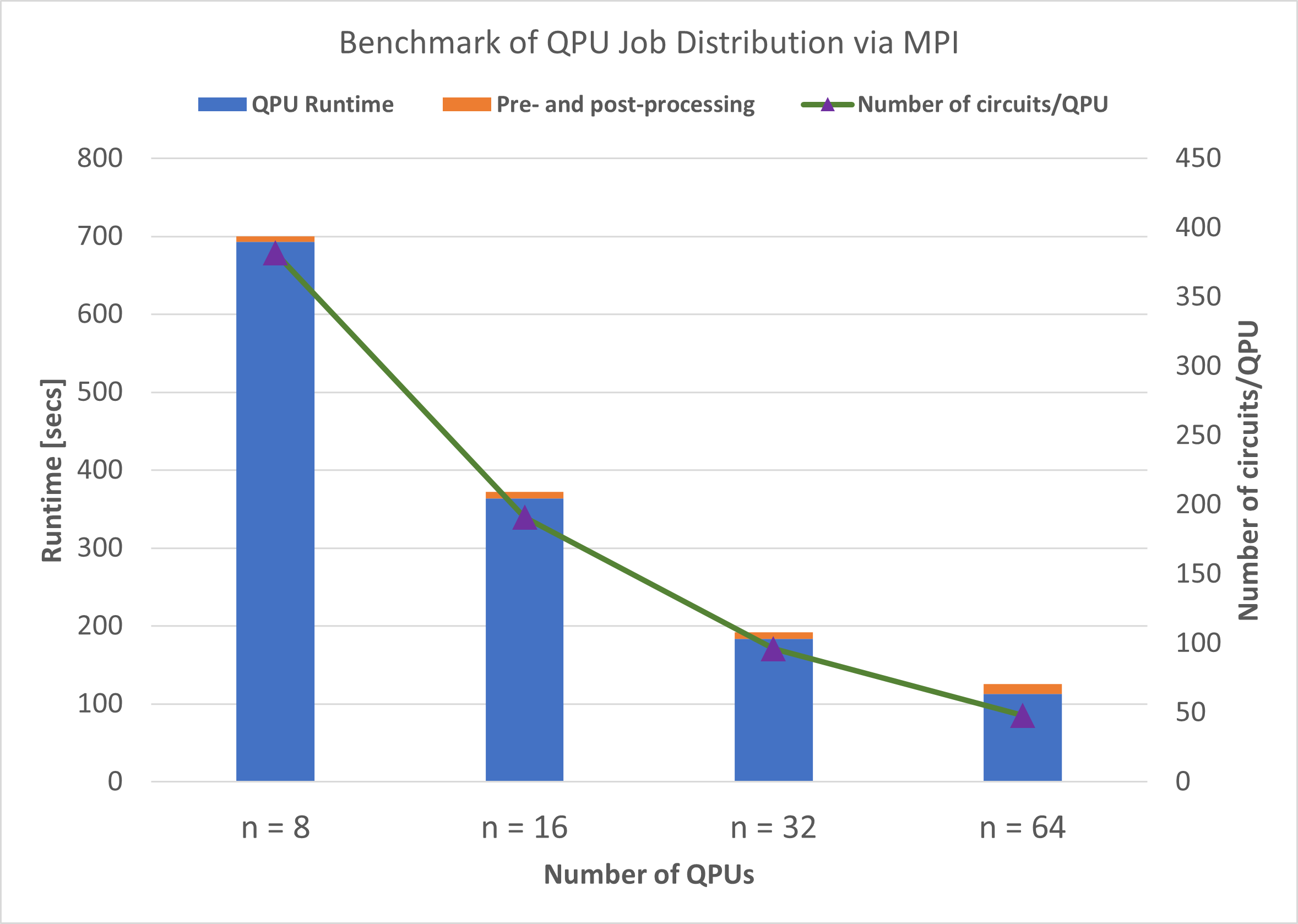}
\caption{Performance scaling with number of (virtual) QPUs for the computation of the ground-state energy of an $H_8$ chain. These calculations make use of a UCCSD ansatz at a fixed set of parameters, with a 16-qubit, 3052-term STO-3G Hamiltonian basis set obtained using the Jordan–Wigner transformation. The SDK reports both the quantum circuit execution time (blue bars) and the time taken for pre-and post-processing (orange bars). We also plot the number of circuits sent to each QPU.}
\label{fig:pawsey_data}
\end{figure}

Fig.~\ref{fig:pawsey_data} shows the time taken to evaluate the ground-state energy of the $H_8$ molecule on different numbers of QPUs (each simulated using a single node of a supercomputing cluster).  The Hamiltonian consists of 3052 terms, requiring many quantum circuit evaluations, each of which includes evaluation of both the ansatz and the change of basis.\footnote{Post-rotations for each term in the Hamiltonian that is not a Pauli $Z$ gate.} It is worth noting that there are optimisation strategies that involve grouping of Pauli terms~\cite[e.g.,][]{izmaylov2019unitary, gokhale2020optimization}, which can help reduce the number of evaluations needed, but for the sake of simplicity we do not employ any grouping strategies for this demonstration.

In this example, the quantum state is prepared by a UCCSD ansatz~\cite{grimsley2019trotterized}, i.e.,
\begin{equation}
    |\psi\rangle = UCCSD |0\rangle.
\end{equation}
at a \emph{fixed} set of variational parameters (i.e.\ no outer classical optimization loop is included in this performance scaling evaluation). We evaluated each of the circuits resulting from the terms in the Hamiltonian using a state-vector-based simulator without statistical noise. 

Data for Fig.~\ref{fig:pawsey_data} result from runs on the Topaz cluster at the Pawsey supercomputing center, with each virtual QPU running on a single Topaz node. Nodes contain 256 GB memory and 2 $\times$ Intel Xeon E5-2680 v4 2.4GHz CPUs, and communicate via a Mellanox InfiniBand interconnect running at 100Gb/s.  MPI is implemented with OpenMPI.

Wall time results in Fig.~\ref{fig:pawsey_data} demonstrate close to ideal scaling when we increase the number of processing nodes (simulated virtual QPUs). More importantly, we observe minimal overhead for pre- and post-processing procedures, such as those associated with MPI scatter and gather steps in the MPI-based virtual QPU implementation of the SDK.

\section{Summary}
\label{sec:Summary}
Thanks to advances in quantum computing hardware, we are fast approaching the regime of quantum utility, where hybrid quantum-classical computers can outperform conventional computers of comparable size, weight and power. A performant and capable software framework that allows for flexible and efficient multi-QPU parallelisation is critical to achieving the scalability and performance required for real-world workloads. Quantum Brilliance, a full-stack quantum computing company, has been developing an SDK incorporating parallelisation strategies targeting different modes of quantum-classical interactions. We have developed a thread-based asynchronous execution model for remotely-hosted QPUs, and a high-performance MPI-based QPU virtualisation system for quantum-accelerated data centers. The SDK enables users to explore hybrid quantum-classical workloads in order to test and benchmark computational utility in their own application domains.    

\section*{Acknowledgement}
This research used resources of the Pawsey Supercomputing Centre, which is supported by the Australian Government under the National Collaborative Research Infrastructure Strategy and related programs through the Department of Education. Daanish Arya, Johannes Kuhlmann and Florian Preis are supported by the BMBF Anwendungsnetzwerkwerk f\"ur das Quantencomputing projects FKZ 13N16233 and FKZ 13N16091.

\bibliographystyle{unsrt}
\bibliography{ref}  
\end{document}